\documentclass{pasj00}

\usepackage{graphicx}
\usepackage{epsfig}
\usepackage{lscape}

\begin{document}

\SetRunningHead{K. Vierdayanti, K.Watarai, and S. Mineshige}{Black hole mass estimation
from X-ray spectral fitting}
\Received{2007/12/06}
\Accepted{2008/02/13}

\title{On Black Hole Mass Estimation from X-ray Spectra of Ultraluminous X-ray Sources}

 \author{%
   Kiki \textsc{Vierdayanti}\altaffilmark{1,}\thanks{Present Affiliation: Department of Astronomy, Kyoto University, Sakyo-ku, Kyoto 606-8502},
   Ken-ya \textsc{Watarai}\altaffilmark{2,3,}\thanks{Research Fellow of the Japan Society for the Promotion of Science},
   and
   Shin \textsc{Mineshige}\altaffilmark{1,$\ast$}}
 \altaffiltext{1}{Yukawa Institute for Theoretical Physics, Kyoto University, Sakyo-ku, Kyoto 606-8502}
 \altaffiltext{2}{Astronomical Institute, Osaka Kyoiku University, Asahigaoka, Kashiwara, Osaka 582-8582}
\altaffiltext{3}{Kanazawa University Fuzoku High School, Heiwa-machi, Kanazawa, Ishikawa 921-8105}

 \email{kiki@yukawa.kyoto-u.ac.jp}


\maketitle

\begin{abstract}
We propose a methodology to derive a black-hole mass for super-critical
accretion flow.
Here, we use the extended disk blackbody (extended DBB) model, a fitting
model in which the effective temperature profile obeys the relation
$T_{\rm eff} \propto r^{-p}$, with
$r$ being the disk radius and $p$ being treated as a fitting parameter.
We first numerically calculate the theoretical flow structure and its spectra
for a given black-hole mass, $M$, and accretion rate, $\dot{M}$.
Through fitting to the theoretical spectra by the extended DBB model,
we can estimate the black-hole mass, $M_{\rm x}$, assuming that the
innermost disk radius is $r_{\rm in}=3r_{\rm g}
(\propto M_{\rm x})$, where $r_{\rm g}$ is the Schwarzschild radius.
We find, however, that the estimated mass deviates from that adopted
in the spectral calculations, $M$, even for low-$\dot{M}$ cases. We
also find that the deviations can be eliminated by introducing a new
correction for the innermost radius. Using this correction, we
calculate mass correction factors, $M/M_{\rm x}$, in the
super-critical regimes for some sets of $M$ and $\dot M$, finding that
a mass correction factor ranges between $M/M_{\rm x} \sim$ 1.2 -- 1.6.
The higher is $\dot{M}$, the larger does the mass correction factor
tend to be.
Since the correction is relatively small, we can safely conclude
that the black holes in ULXs which Vierdayanti et al. (2006, PASJ, 58, 915)
analyzed are stellar-mass black holes with the mass being $< 100M_{\odot}$.
\end{abstract}

\section{Introduction}
The most intriguing question regarding the ultraluminous X-ray sources (ULXs) 
at present would be how massive is the black hole in each ULX. 

Known to be very bright extragalactic X-ray sources
($L_{\rm x}\sim 10^{39 - 41}{\rm erg} {\rm s}^{-1}$), brighter than
the ever-known Galactic X-ray sources, ULXs are good candidate to prove
the existence of the hypothetical new class of black holes,
the intermediate mass black holes (IMBHs), whose masses range over
$10^{2}$ -- $10^{4}$ $M_{\odot}$ (e.g. Colbert \& Mushotzky 1999; \cite{25}). 
If we simply assume $L_{\rm x} \leq L_{\rm E}$, where $L_{\rm E}$ is
the Eddington luminosity, $M \geq 10^{3} M_{\odot}$ for $L_{\rm x}
\sim 10^{41}$ erg s$^{-1}$.
This provides strong support to the IMBH hypothesis.

The success in fitting the spectra of several ULXs with a multicolor disk 
blackbody (DBB) and a power-law (PL) model further supports the idea of
IMBHs for the ULXs, since the cool inner disk temperatures and the
large innermost radii, obtained through model fitting, suggest a black
hole mass within the IMBH range (Miller et al. 2003,2004; \cite{9}, \cite{36}).
Moreover, the explanation behind the IMBH notion for ULXs is based on the 
standard accretion disk model (Shakura \& Sunyaev 1973),
a well-established model for high-energy release mechanism through gas 
accretion flows around a compact object.

On the other hand, some attempts have been made to challenge the IMBH 
interpretation of ULXs (King et al. 2001). Several models have been proposed 
to allow stellar-mass black holes to power ULXs. In fact, such attempts 
are very reasonable since the existence of stellar-mass black holes is well
founded while IMBH formation has not been well understood
(e.g. Madhusudhan et al. 2006).

Amongst several models proposed, a model of super-critical accretion flow onto 
a stellar-mass black hole has been quite favorable recently (e.g., \cite{48};
\cite{33}; \cite{53}; \cite{56}).
One attractive model supporting a super-critical accretion flow onto a
stellar-mass black hole is the slim disk model (\cite{46}; \cite{20},
\cite{14}; \cite{33}) introduced in the late 1980s by the Warsaw and the
Kyoto groups (\cite{3}, 1989, see \cite{19} for a review).
Its application to some Galactic black holes was investigated by Watarai
et al. (2000) in their attempts to explain the nature of some bright X-ray
source binaries. Later on, Watarai et al. (2001) also proposed the slim disk
model for ULXs.

The black-hole mass mentioned above was estimated from an X-ray spectral
fitting in
which the innermost radius, $r_{\rm in}$, was assumed to coincide with 
that of the last stable circular orbit, $r_{\rm ms}$ ($=3r_{\rm g}$ for a
non-rotating black hole, where $r_{\rm g}$ [$\equiv 2GM/c^2$] is the
Schwarzschild radius).
This assumption has been proven to be good for the cases of 
Galactic black hole binaries (e.g., \cite{72}; Tanaka \& Shibazaki 1996;
\cite{12}; \cite{1010}).

However, numerical calculations have shown that as mass accretion rate
increases, the flow inside $3r_{\rm g}$ might produce significant emission
(\cite{46}).
The heat generated near $3r_{\rm g}$ is trapped by the inward material flows,
and is emitted inside $3r_{\rm g}$. Some amount of the trapped radiation going
closer to the black hole is swallowed.
Therefore, since the situation is more complicated in high mass accretion 
rate systems, we cannot simply relate the innermost radius from the
fitting with the last stable circular orbit.
Indeed, Watarai and Mineshige (2003a) have stressed that the observed value
of the inner edge of the disk is more like a radiation edge,
outside of which substantial emission is produced.
In this work, we investigated the assumption $r_{\rm in}=3r_{\rm g}$ for 
the super-critical accretion flows around a non-rotating black hole and
calculate corrections to make more accurate black-hole mass estimations.

The plan of this paper is as follows:

We describe the disk structure and spectrum calculations in the first
two subsections in section 2, and the fitting model, 
together with the conventional method to derive black hole mass from the 
fitting in the remaining subsection.

In section 3, we present our fitting analysis and results,
while focusing on some sets of parameters.

Section 4 is devoted to a discussion, and finally section 5 concludes the paper.

\section{Methodology}
To avoid confusion, we will first describe the overall picture of our 
methodology which is also summarized in table 1.
The detail steps in the fitting are explained in another section
(section 3).

The main steps of our methodology are as follows:
We first calculate the disk structure for various values of black-hole mass,
$M$, and the mass-accretion rate, $\dot{M}$ (subsection 2.1). 
Next, we use the outputs from the disk structure calculation to obtain 
the spectra (subsection 2.2).
We follow the calculation by Watarai et al. (2005) for both steps.
We then fit those spectra with the extended DBB model and derive the black hole
mass, $M_{\rm x}$ from value of $r_{\rm in}$ obtained from the fitting
(subsection 2.3). 
That is, we assume $r_{\rm in}=3r_{\rm g}$ as that of the DBB model.
Finally, we obtain the correction factor, $M/M_{\rm x}$, the ratio of the
actual black hole mass to the derived black hole mass (subsection 3.2).

\begin{center}
\begin{longtable}{lllll}
  \caption{The overall picture of our methodology.
}\label{tab:first}
\hline \hline
  Step & Calculation & Input & Output & Section  \\
  \hline
\endhead
 \hline
\endfoot
 \hline
\endlastfoot
 1   & Disk structure & $M$, $\dot{M}$ & $T_{\rm eff}(r)$, $H(r)$, $v_{r}(r)$, $v_{\varphi}(r)$ & $2.1$ \\
 2   & Spectrum & $T_{\rm eff}(r)$, $H(r)$, $v_{r}(r)$, $v_{\varphi}(r)$ & $S_{\nu}$ & 2.2 \\
 3   & Fitting & $S_{\nu}$ & $T_{\rm in}$, $r_{\rm in}$, $p$ & 2.3 \\
 4   & Correction factor & $r_{\rm in}$, $M$ & $M/M_{\rm x}$ & 3.2 \\
   \hline \hline 
\end{longtable}
\end{center}

\subsection{Disk Structure Calculation}
The standard accretion-disk model has been very successful in describing 
optically thick flow structure as long as the mass accretion rate of the 
systems in question is less than the critical mass-accretion rate
\footnote{Some papers use
$\dot{M}_{\rm crit} \equiv L_{\rm E}/(\eta c^{2})$ with $\eta$,
the energy conversion efficiency, being a constant. However,
this may cause confusion since $\eta$ is not constant in the slim disk model.},
$\dot{M} \leq \dot{M}_{\rm crit}\equiv L_{\rm E}/c^{2}$.

It is possible, however, that the mass-accretion rate of a system 
exceeds the critical mass-accretion rate in disk accretion because of
anisotropic radiation field (Ohsuga et al. 2005).  
A disk in which its accretion rate exceeds the critical value is called a 
super-critical accretion disk. 

In this section we describe the calculation of the structure of such a disk
(e.g., \cite{46}, 2005; see reviews by \cite{19}).
The equations are written by using cylindrical coordinates
$(r,\varphi,z)$. 
A non-rotating black hole is assumed as the central object and a
pseudo-Newtonian potential (Paczy\'{n}ski \& Wiita 1980) is adopted to
model the
general relativistic effects in the gravitational field of the black hole,
\begin{equation}
  \psi=-\frac{GM}{R-r_{\rm g}},
 \label{118}
 \end{equation}  
where $R=\sqrt{r^{2}+z^{2}}$.

The pressure and density are related to each other by polytropic relation
(H$\bar{\rm o}$shi 1977), $p\propto\rho^{1+\frac{1}{N}}$, in the vertical
direction, where $N$ is the polytropic index.
The vertically integrated density, $\Sigma$, and pressure, $\Pi$, are then
\begin{equation}
  \Sigma \equiv \int_{-H}^{H} \rho dz=\int_{-H}^{H} \rho_{0} \left( 1-\frac{z^{2}}{H^{2}}\right)^{N} dz=2\rho_{0}I_{N}H,
 \label{119}
 \end{equation}   
\begin{equation}
  \Pi \equiv \int_{-H}^{H} p dz=\int_{-H}^{H} p \left( 1-\frac{z^{2}}{H^{2}}\right)^{N+1} dz=2p_{0}I_{N+1}H,
 \label{120}
 \end{equation}   
where 
\begin{equation}
  I_{N}=\frac{(2^{N}N!)^{2}}{(2N+1)!},
 \label{121}
 \end{equation}  
and $H$ is the scale height.

Assuming hydrostatic balance in the vertical direction, we have
\begin{equation}
  \Omega_{\rm K}^{2}H^{2}=2(N+1)\frac{p_{0}}{\rho_{0}},
 \label{122}
 \end{equation}  
where $\Omega_{\rm K}$ is the Keplerian angular velocity of the disk rotation
in the pseudo-Newtonian potential, which is defined by
\begin{equation}
  \Omega_{\rm K}=\left(\frac{\partial{\psi}}{r\partial{r}}\right)^{1/2}\Bigg|_{z=0}=\frac{r}{r-r_{g}}\left(\frac{GM}{r^{3}}\right)^{1/2}.
 \label{123}
 \end{equation} 
Note that in terms of $\Pi$ and $\Sigma$, the hydrostatic balance in the
vertical direction can be written as
\begin{equation}
  \Omega_{\rm K}^{2}H^{2}=(2N+3)\frac{\Pi}{\Sigma}.
 \label{124}
 \end{equation}
Regarding the polytropic index, we set $N=3$ for the whole disk calculation.

The other basic equations are obtained by integration in the vertical
direction under the assumption that the radial velocity, $v_{r}$, and the
specific angular momentum, $\ell$ ($=rv_{\varphi}=r^{2}\Omega)$, do not
depend on the vertical coordinate.

The continuity equation gives
\begin{equation}
  -2\pi r \Sigma v_{r}=\dot{M}={\rm constant},
 \label{125}
 \end{equation}  
where $\dot{M}$ is the mass accretion rate.

The radial component of the momentum equation can be written as
\begin{equation}
  v_{r}\frac{dv_{r}}{dr}+\frac{1}{\Sigma}\frac{d\Pi}{dr}=\frac{\ell^{2}-\ell_{\rm K}^{2}}{r^{3}}-\frac{\Pi}{\Sigma}\frac{d\ln\Omega_{\rm K}}{dr},
 \label{126}
 \end{equation}  
where $\ell_{\rm K}$ is the Keplerian angular momentum, defined by
$\ell_{\rm K}=r^{2}\Omega_{\rm K}$. The last term of the above equation,
$\frac{\Pi}{\Sigma}\frac{d\ln\Omega_{\rm K}}{dr}$, is a correction term
resulting from the fact that the radial component of the gravitational force
changes with height (Matsumoto et al. 1984).

The angular-momentum balance is obtained from the $\varphi$-component of
momentum equation, and can be written as
\begin{equation}
  \dot{M}(\ell-\ell_{\rm in})=-2\pi r^{2}T_{r\varphi},
 \label{127}
 \end{equation} 
where $\ell_{\rm in}$ is an integration constant, which represents the
specific angular momentum finally swallowed by the black hole, and
$T_{r\varphi}$ is the vertical integration of $t_{r\varphi}$, that is
\begin{equation}
  T_{r\varphi}\equiv \int_{-H}^{H}t_{r\varphi}dz=-\alpha \Pi,
 \label{128}
 \end{equation} 
with $t_{r\varphi}=-\alpha p$ as in the standard Shakura-Sunyaev prescription.

The energy equation can be written as
\begin{equation}
  Q_{\rm adv}^{-}=Q_{\rm vis}^{+}-Q_{\rm rad}^{-},
 \label{129}
 \end{equation} 
where $Q_{\rm adv}^{-}$ is the advective cooling, defined by
\begin{equation}
  Q_{\rm adv}^{-}=\int_{-H}^{H}q_{\rm adv}^{-}dz=\frac{9}{8} v_{r} \Sigma T_{\rm c} \frac{ds}{dr},
 \label{130}
 \end{equation}
where $s$ is the specific entropy on the equatorial plane.
The viscous energy generation rate is given by
\begin{equation}
  Q_{\rm vis}^{+}=\int_{-H}^{H}q_{\rm vis}^{+}dz=rT_{r\varphi}\frac{d\Omega}{dr},
 \label{131}
 \end{equation}
while the radiative cooling rate is given by
\begin{equation}
  Q_{\rm rad}^{-}=\int_{-H}^{H}q_{\rm rad}^{-}dz = 2F,
 \label{132}
 \end{equation}
where $F$ is the radiative flux per unit surface area on the disk surface,
given by
\begin{equation}
  F=\frac{8acT_{\rm c}^{4}}{3\tau},
 \label{133}
 \end{equation}
where $a$ is radiative constant, $T_{\rm c}$ is the temperature on the
equatorial plane, and $\tau$ is given by
\begin{equation}
  \tau=\bar{\kappa}\Sigma=(\kappa_{\rm es}+\kappa_{\rm ff})\Sigma.
 \label{134}
 \end{equation}

The equation of state is given by
\begin{equation}
  \Pi=\Pi_{\rm gas}+\Pi_{\rm rad}=\frac{k_{\rm B}}{\bar{\mu}m_{\rm H}}\frac{I_{4}}{I_{3}}\Sigma T_{\rm c}+\Pi_{\rm rad},
 \label{135}
 \end{equation}
where $\Pi_{\rm rad}$ is defined by
\begin{equation}
  \Pi_{\rm rad}\equiv \int_{-H}^{H}\frac{1}{3} a T_{\rm c}^{4}\left(1-\frac{z^{2}}{H^{2}}\right)^{4}dz.
 \label{136}
 \end{equation}

We first specify the values of $M$, $\alpha$, $\dot{M}$, and an initial guess of
$\ell_{\rm in}$.
The outer boundary conditions are imposed at $r=1\times 10^{4}r_{\rm g}$, 
where all physical quantities are taken as those of the standard disk.
The inner calculation boundary is taken at $r\sim 2.0 r_{\rm g}$. 
Note that it is important to solve the equations, even inside the marginally
stable last circular orbit at $3 r_{\rm g}$, since substantial amount of
matter still exists there.
In fact, the emission from this region makes a rather important contribution
when $\dot{M}\geq 100\dot{M}_{\rm crit}$.
Also note that it is possible to calculate the disk structure inside
$2r_{\rm g}$, but in this region the disk in fact becomes optically thin and
thus the blackbody assumption made in calculating the spectrum may no longer
be valid.
Further, the contribution from the innermost part at $r < 2r_{\rm g}$ is small
because of a photon redshift.
Thus, we left this region for our future studies. 

The solution should satisfy the regularity condition at the transonic radius
located between (2 -- 3) $r_{\rm g}$ and the free boundary conditions at the
inner edge.
The basic equations are first radially integrated for an a priori given
$\ell_{\rm in}$. 
When $\ell_{\rm in}$ is larger (or smaller) than the correct value, the
radial velocity vanishes (diverges) before $r_{\rm g}$.
Iteration of $\ell_{\rm in}$ is therefore needed to find an appropriate value
for which a transonic solution is obtained.    
In other words, the value of $\ell_{\rm in}$ is determined uniquely by the
transonic condition.

\subsection{Spectrum Calculation}
In calculating the spectrum, we apply a calculation method proposed by
Watarai et al. (2005). 
They adopted the 'Ray-Tracing method' (Luminet 1979; Fukue \& Yokoyama 1988)
in calculating the spectra of a disk around a Schwarzschild black hole.
Before reaching the observer, a photon emitted from some point on the disk
travels along the null geodesics.
In the Schwarzschild space-time, the photon's trajectory is determined by
\begin{equation}
  \frac{d^{2}}{d\chi^{2}} \left(\frac{1}{r}\right)+ \frac{1}{r} = \frac{3 r_{\rm g}}{2r^{2}},
 \label{158}
 \end{equation}
where $r$ is the distance from the
center (black hole), and $\chi$ is the angle measured from the direction of
the observer (see reviews by \cite{19}).
The trajectories will then satisfy the following differential equation:
\begin{equation}
  \left[\frac{d}{d\chi} \left(\frac{1}{r}\right)\right]^{2}+ \frac{1}{r^{2}} \left(1-\frac{r_{\rm g}}{r}\right) = \frac{1}{b^{2}},
 \label{159}
 \end{equation}
where $b$ is the impact parameter.

The light ray is calculated from an observer's coordinate to the surface of
the disk by the 'Runge-Kutta-Gill Method' and the scale-height is defined
as the disk surface of the last scattering.
The numerical integration is stopped when the ray arrives at the surface of
the disk, and the final arrival point is interpolated from the calculated data
of the disk model. 
The obtained quantities, such as temperature, velocity field, and redshift on
the disk surface, are transformed to the observer's values, which are used to
calculate the observed spectrum.

The observed spectra were calculated by using the Lorentz invariant relation,
\begin{equation}
  I_{\nu_{\rm obs}}= \left(\frac{\nu_{\rm obs}}{\nu_{\rm em}}\right)^{3} I_{\nu_{\rm em}} \approx \frac {1}{(1+z)^{3}} B_{\nu_{\rm em}}(T_{\rm eff}),
 \label{160}
 \end{equation}
where $I_{\nu_{\rm em}}$ and $\nu_{\rm em}$ are the intensity and frequency in
the disk co-moving frame, respectively.
Some simplifications are also made here, that the blackbody radiation,
$B_{\nu}(T_{\rm eff})$, is assumed in order to calculate the spectra, and the
temperature on the disk surface is adopted as the effective temperature.
The thermal Comptonization process, which may become important in the innermost
region, is approximated by introducing a spectral hardening factor, $\kappa$.
We set $\kappa=1.7$ throughout the calculation (see discussion in
subsection 4.2).
The local flux is calculated by using a diluted blackbody approximation,
\begin{equation}
  F_{\nu}^{\rm db}= \frac{1}{\kappa^{4}} \pi B_{\nu}(\kappa T_{\rm eff}).
 \label{161}
 \end{equation} 
The redshift factor, $(1+z)$, which consists of a gravitational redshift
part and the Doppler parts, is given by,
\begin{equation}
  1+z = \frac {E_{\rm em}}{E_{\rm obs}} = L^{-1}\gamma D^{-1},
 \label{161}
 \end{equation}
where $L$, $\gamma$, and $D$ are the lapse functions representing the
gravitational redshift, Lorentz factor, and Doppler factor, respectively
(Kato et al. 2008).

According to Watarai et al. (2005), the inclination-angle dependence of the 
accretion disk spectrum is composed of the following three factors:
\begin{enumerate}
\item The projection effect, which is the change of the effective area with
the change of the inclination angle.
\label{1}
\item The effect of Doppler beaming towards the observer's direction, which
causes asymmetry in the flux distribution with respect to the rotation axis.
\label{2}
\item The self-occultation effect (Fukue 2000), that is, the central part of
the disk is obscured by the cooler outer part of the disk. This effect 
is significant for high-inclination systems.
\label{3} 
\end{enumerate}
Note that self-irradiation (Cunningham 1975) was not considered in the present
study.

The spectrum calculation done by Watarai et al. (2005) was not intended
for calculating the spectrum over a large area of the disk. We made a
simple modification to make the calculation for larger disk area
possible with reasonable calculation time. This is necessary since the
spectra of the high mass-accretion rate disks extend into the soft
X-ray region (Szuszkiewicz et al. 1996). To do so, we divided the disk
by following $\sqrt{r}$ pattern. By doing so, we can keep the detail of the
innermost part without losing those of the outermost part too much.

\subsection{Fitting Model}
In this subsection we switch to a description of the fitting model and the
conventional method to derive the black hole mass.
\subsubsection{Extended disk blackbody model}
The fitting model that we describe in this section, and is the main
tool used in this work, is the extended disk blackbody (extended DBB) model.
This X-ray spectral fitting model was originally proposed by Mineshige et al.
(1994)
to investigate whether the temperature gradient found from the observations of
Nova Muscae 1991 could deviate from the canonical value of the standard disk,
0.75 ($T_{\rm eff} \propto r^{-3/4}$).

The name {\lq extended DBB\rq} is chosen, because it is a natural extension of
the DBB model
\footnote{In some papers it is called the \lq $p$-free\rq model, but we
do not use it, since this is a confusing terminology.}.
It becomes a generalization and an extension of the DBB model, since the
temperature gradient is now treated as one of the fitting parameters.
The temperature profile, hence, becomes,
\begin{equation}
  T_{\rm eff} = T_{\rm in}\left(\frac{r}{r_{\rm in}}\right)^{-p},
 \label{162}
 \end{equation} 
where $T_{\rm in}$ and $r_{\rm in}$ have the same interpretation as those 
in the DBB model. 

By using self-similar solutions, Watarai and Fukue (1999) found that the
temperature distribution of the slim disk is somewhat flatter than that of
the standard disk, that is 
$T_{\rm eff} \propto r^{-1/2}$ (see also Wang \& Zhou 1999). 
Therefore, the temperature profile of the standard disk is proportional
to $r^{-3/4}$, while for the case of slim disk, $r^{-1/2}$.

Generally speaking, hence, by using the extended DBB model, $r^{-p}$, for the 
fitting, we can distinguish the case where the standard disk is more plausible 
than the slim disk, and vice versa, from the spectrum profile.

\begin{figure*}
  \begin{center}
\centerline{\epsfig{file=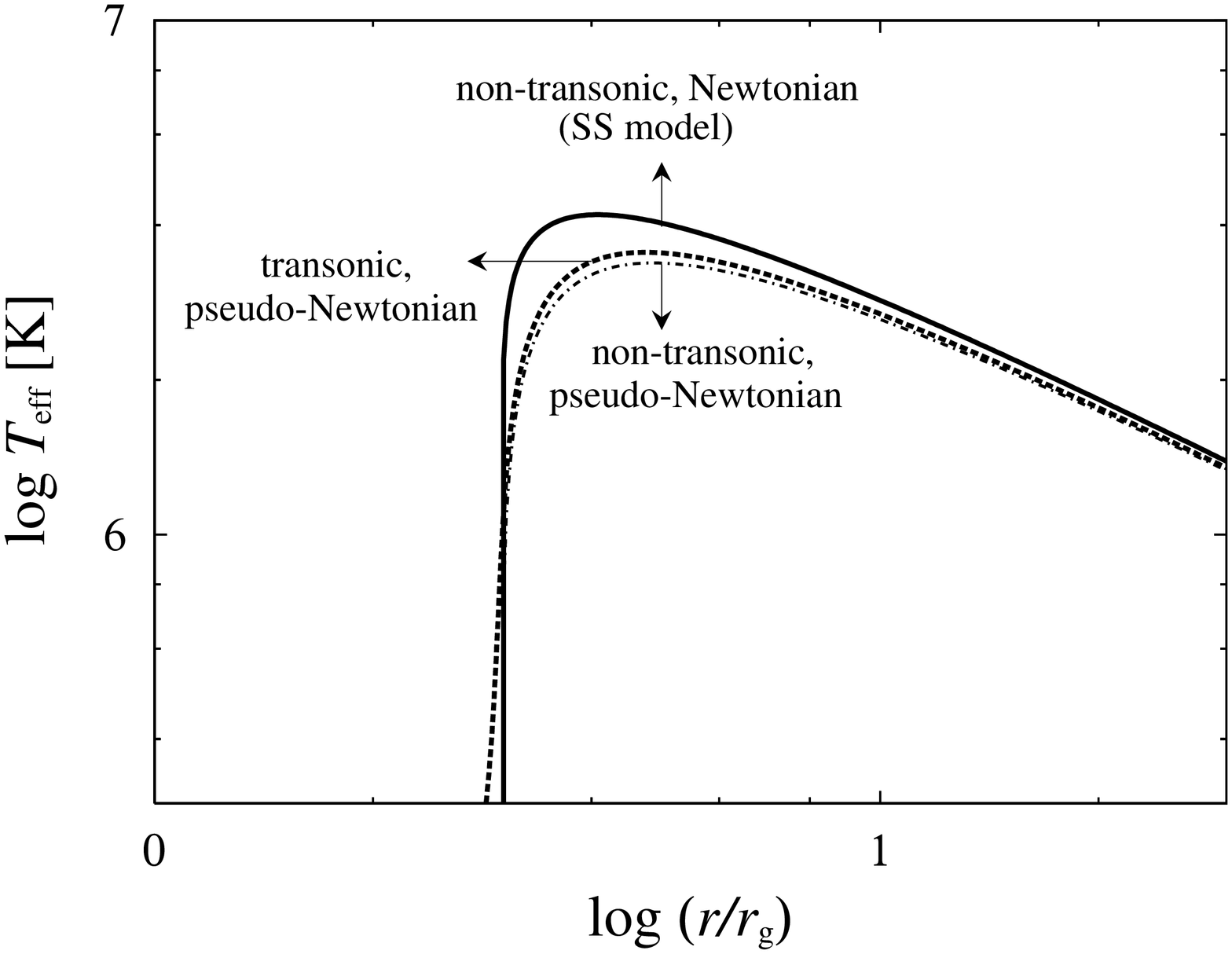,width=6.5cm,angle=270}\epsfig{file=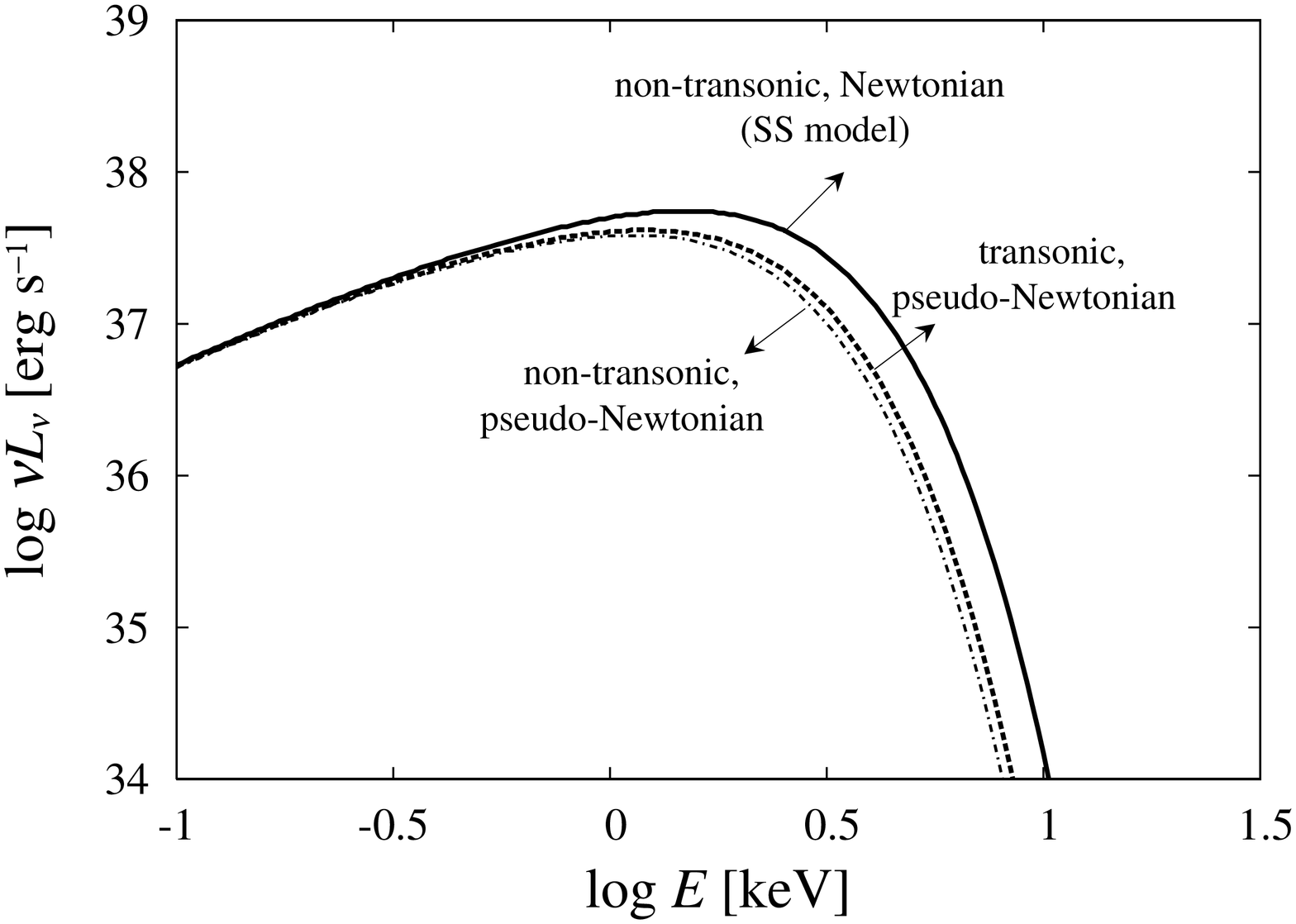,width=6.5cm,angle=270}}
  \end{center}
  \caption{Comparison between the standard disk model (Shakura and Sunyaev 1973; analytical Newtonian non-transonic solution) in solid lines and the numerical model (pseudo-Newtonian, transonic solution) at low $\dot{m}$ ($\dot{m}=1$), in dashed lines. Left panel: effective temperature profiles. Right panel: Spectra. We also plot analytical pseudo-Newtonian non-transonic solution in thin dot-dashed lines.}\label{Figure:3}
\end{figure*}
\subsubsection{Conventional methodology to derive black-hole mass}
When we obtain a good fit to the observed spectra by the DBB model,
we can easily estimate the black-hole mass and the Eddington ratio,
$L/L_{\rm E}$ (with $L$ and $L_{\rm E}$ being disk luminosity
and the Eddington luminosity, respectively), based on the standard disk
theory.

The basic methodology is summarized in Makishima et al. (2000).

We write the bolometric luminosity of an optically thick accretion disk as
 \begin{equation}
  L_{\rm bol} = 2 \pi D^2 f_{\rm bol} (\cos{i})^{-1},
 \label{1}
 \end{equation}
with $f_{\rm bol}$ being the bolometric flux, $i$ being the inclination of the
disk ($i=0$ corresponds to face-on geometry) and {\it D} is the distance.
This $L_{\rm bol}$ is related to the maximum disk color temperature, 
$T_{\rm in}$, and the innermost disk radius, $r_{\rm in}$, as

 \begin{equation}
  L_{\rm bol} = 4 \pi (r_{\rm in}/\xi)^2 \sigma (T_{\rm in}/\kappa)^4,
 \label{2}
 \end{equation} 
where $\kappa \sim 1.7$ (Shimura \& Takahara 1995) is the ratio of the color
temperature to the effective temperature, or spectral hardening factor, and
$\xi = 0.412$ is correction factor reflecting the fact that $T_{\rm in}$
occurs at somewhat larger than $r_{\rm in}$.

Hence, the innermost disk radius is 

 \begin{equation}
  r_{\rm in} = \xi \kappa^2 \sqrt {\frac{L_{\rm bol}}{4\pi\sigma T_{\rm in}^4}}.
 \label{3}
 \end{equation}
We may identify $r_{\rm in}$ with the radius of the last stable Keplerian orbit. Thus, we may in general write

 \begin{equation}
  r_{\rm in} =3\beta r_{\rm g} = 8.86\beta (\frac{M}{M_{\odot}}) \ {\rm km} \ \ \
{\rm with}\ \ \ \ \frac{1}{6} < \beta \leq 1 
 \label{4}
 \end{equation} 
by which we can determine the black hole mass.
Note that $\beta$ depends on the black hole spin and $\beta = 1/6$ for an
extremely rotating Kerr black hole, and $\beta=1$ for a Schwarzschild black
hole. In the present study, we focus on the cases of the Schwarzschild black
hole ($\beta = 1$).

\subsubsection{Problem with the Conventional Methodology}
The extended DBB model is a phenomenological model which was proposed to 
investigate a possible deviation of the temperature gradient from 0.75, 
the value predicted by the theory of the standard disk.
It turned out that this model can also be useful to discriminate the
slim disks from the standard disks, as explained in the previous part.

However, physical interpretation of the fitting results by using the extended
DBB model is far from being clear, unlike in the case of DBB model. 
We need a methodology to make a physical interpretation of the extended DBB
model in the light of the super-critical accretion-disk theory.

In the case of ULXs, Vierdayanti et al. (2006), for example, have tried to
make a physical interpretation of the fitting results from the extended DBB
fitting model. 
They adopted a method that is widely used for the DBB fitting model which 
corresponds to the standard disk theory to derive black hole masses from the
fitting results.
The fitting shows the tendency that the observational data of the analyzed 
ULXs are more likely to be fitted with the slim disk model.
We realized that the adoption of the conventional method used for the DBB model 
is not fully appropriate, since the spectral shapes of the slim disk is
different from those of the standard disk. 
Therefore, the adoption of the conventional method is inadequate for estimating 
the black hole mass. 
In addition, a number of relativistic effects as well as self-occultation
effect certainly affect the spectra, and thus the fitting results. 
The implication of such relativistic cases are not obvious, and have been
poorly investigated so far.
In conclusion, some correction should be introduced to permit this adoption.
To be precise, in the following section we will try to find a mass correction
factor, defined as $m/m_{\rm x}$, where $m$ ($\equiv M/M_{\odot}$) and
$m_{\rm x}$ ($\equiv M_{\rm x}/M_{\odot}$) are the original black-hole mass
and the black-hole mass obtained from the fitting, respectively. 

\begin{center}
\begin{longtable}{lllll}
  \caption{The fitting steps (see text for definition of $\xi^{\rm trans}$, and $M/M_{\rm x}$).
}\label{tab:first}
\hline \hline
  Step & Fitted Model & $\dot{m}$-value & Corrections & Section  \\
  \hline
\endhead
 \hline
\endfoot
 \hline
\endlastfoot
 0   & Shakura-Sunyaev model & 1,10 & $\xi=0.412$ &  3.1 \\
 1    & Numerical transonic model & 1,10,32,100,320,1000 & $\xi^{\rm trans}=0.353$, $M/M_{\rm x}$ (for high-$\dot{M}$) & 3.1 and 3.2 \\
   \hline \hline 
\end{longtable}
\end{center}
\begin{figure}
  \begin{center}
\centerline{\epsfig{file=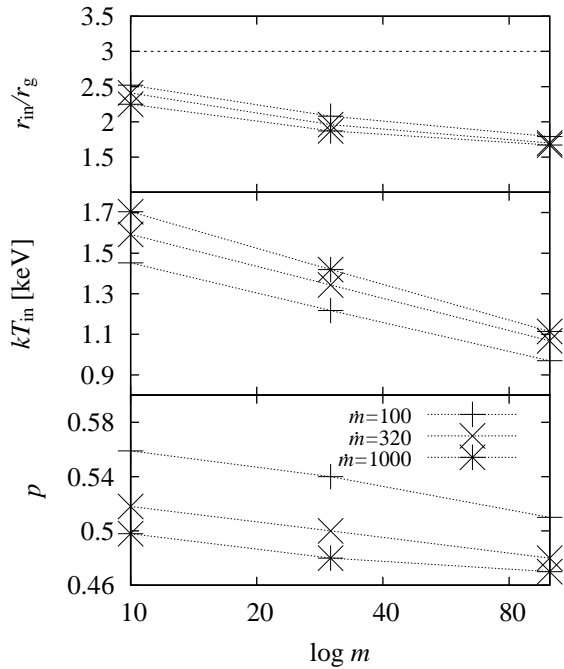,width=9.0cm,angle=270}}
  \end{center}
  \caption{Fitting results: $p$ (bottom panel), $kT_{\rm in}$ (middle panel), and $r_{\rm in}/r_{\rm g}$ (top panel) with $\dot{m}=100$, $320$, and $1000$ and $i=0$ for various $m$.}\label{Figure:3}
\end{figure}
%
\section{Fitting Analysis and Results}
We fit the theoretical spectra of the slim disk (Watarai et al. 2005) with the 
extended DBB model. In the theoretical spectrum calculation, there are three 
parameters that should be determined in the disk structure calculation, whose 
results are used to calculate the spectra. 
They are, the black-hole mass, $m(\equiv M/M_{\odot})$, the mass-accretion
rate, $\dot{m}(\equiv \dot{M} c^{2}/L_{\rm E})$, and the viscosity 
parameter, $\alpha$. Regarding the viscosity parameter, in this study we 
only focus on $\alpha=0.01$. Another parameter is needed in the spectrum 
calculation, that is the inclination angle of the disk or the viewing angle 
of the observer.
On the other hand, there are three fitting parameters in the extended 
DBB model, $T_{\rm in}$, $r_{\rm in}$, and $p$.
We divide our fitting procedure into two important steps, which are summarized
in table 2.

To relate $r_{\rm in}$ with the radius of the inner edge of the disk,
we need a correction, $\xi$. However, we found numerically that the
$\xi$-value does not coincide with the analytically expected one,
$\xi=0.412$. Hence, we firstly fit two disk models, analytical standard disk
model and numerical disk model for low $\dot{m}$-case, with the extended DBB
model to find the basic corrections, which will be explained in the following
subsection. The details are given in subsection 3.1 

Secondly, after we found the basic corrections we applied them to the high
$\dot{m}$-case.
This step is described in subsection 3.2.

\subsection{Basic Corrections}
As a first step of our fitting, we elucidate some basic corrections that
were set aside in some previous studies.
We call them basic corrections, since they emerged from the disk structure
calculation before we calculated the spectra. 

In the fitting with the extended DBB model, which is just an extension of the
DBB model, we expect the same results for both analytical standard disk model
and numerical disk model for the case of moderately low $\dot{m}$,
$\dot{m} \le 1$.  
This is because theoretically, the disk should behave like a standard disk when
$\dot{m}$ is moderately low.
On the other hand, it should deviate from that of the standard disk
when $\dot{m}$ is moderately high, since the assumptions made for the standard
disk case will break down.

Interestingly, our fitting results show some discrepancies even for a
moderately small $\dot{m}$ case.
We studied some previous works and come to a conclusion that a
possible source of discrepancies is q small deviation of the 
numerical disk model from that of the standard disk.
(cf. \cite{1414}; \cite{90}: Kubota 2001).
In order to prove it, we compared the numerical transonic flow model in
a pseudo-Newtonian potential with the analytical Shakura and Sunyaev
(SS) model (in which the transonic nature of the flow is not considered) in
a Newtonian potential.

The left panel of figure 1 shows the effective temperature profiles of both
models.
The solid lines represent the analytical result of Shakura-Sunyaev model,
while the dashed lines represent that of the transonic model 
for $m=10$ and $\dot{m}=1$.
We can see that even at low $\dot{m}$, a discrepancy exists, especially in the
innermost region, where the transonic solution plays its role and the effect
of different potential forms is more significant.
We then calculate the spectra of both models by using the spectrum calculation
code of Watarai et al. (2005). However, here and only when it is stated, we
neglect the relativistic Doppler and light bending effects for simplicity.
The results are shown in the right panel of figure 1.
In addition, we also plot analytical pseudo-Newtonian
non-transonic solution as a comparison (thin dot-dashed lines in figure 1).
As shown in figure 1, the effect of the difference in the choice of
potential form is more prominent compared to the effect of
transonic/non-transonic solution.

The discrepancy in the temperature profiles shows that we need a correction
factor in addition to $\kappa$ for the inner temperature.
In order to find the required correction factor, we calculate the spectrum of
those two disk structure models without considering any spectral hardening
effect, i.e., we set $\kappa=1$ and compare the values of $T_{\rm in}$.
We found that $T_{\rm in}$ of the numerical model is by a factor of
$\sim 1.2$ lower than that of the standard disk.
Thus, in fitting the numerical disk model with the extended DBB fitting model,
the inner temperature correction $T_{\rm in}^{\rm SS}/T_{\rm in} = 1.2$ is
required in addition to $\kappa$, where $T_{\rm in}^{\rm SS}$ in the inner
temperature of the Shakura-Sunyaev model.
Therefore, it is necessary to define
$\kappa^{\rm real} \equiv \kappa^{\rm trans} \kappa$,
as a real correction for the inner temperature, where
$\kappa^{\rm trans}=1/1.2$ and $\kappa \sim 1.7$ for Compton scattering.
 
We found that the innermost radius, $r_{\rm in}$, also deviates from
that of the standard disk.
Note that in calculating $r_{\rm in}$, we have used the bolometric luminosity,
and thus the luminosity does not depend on the fitting energy range.
We decided to handle this deviation separately from that of the temperature.
The reason is to keep the black-hole mass estimation method simple.
That is, we can simply obtain $r_{\rm in}$ from the $T_{\rm in}$
(without correcting it first), and just add the correction at the end of
the procedure, that is after we obtain the estimated black hole mass.
The correction for $T_{\rm in}$ becomes important when we want to estimate the mass accretion rate which is not the focus of our present study.

By using the uncorrected $T_{\rm in}$ in the fitting,
the deviation in the innermost radius can be eliminated by changing the
correction factor, $\xi$, introduced earlier (see subsubsection 2.3.2).
In our present study, we found that the new value of $\xi$,
$\xi^{\rm trans}=0.353$ is
more appropriate instead of $\xi=0.412$ which is commonly used
(cf. \cite{1717}).  
We summarize the fitting results in table 3.

We notice that the $p$-value obtained by the fitting is not close to $p=0.75$
but significantly lower, $p<0.64$.
This reflects a bit flatter temperature profile of the transonic solution
(see figure 1)

\begin{table}
  \caption{Fitting results with the extended DBB model for $\alpha=0.01$ and $i=0$. The unit for $kT_{\rm in}$ is keV and energy range for fitting is 0.3 -- 10 keV. The luminosity is calculated for all energy range (not only 0.3 -- 10 keV).
}\label{tab:first}
  \begin{center}
    \begin{tabular}{rrrrrrr} 
\hline \hline
 $m$ & $\dot{m}$ & $p$ & $kT_{\rm in}$ & $r_{\rm in}/r_{\rm g}$ & $\kappa^{\rm trans}$ & $\xi^{\rm trans}$ \\
  \hline
 $10$ & $1$   & $0.637$ & $0.53$ & $2.96$ & $1/1.19$ & $0.353$\\
 $10$ & $10$  & $0.634$ & $0.96$ & $2.79$ & $1/1.15$ & $0.353$\\
  \hline
 $30$ & $1$   & $0.639$ & $0.40$ & $2.97$ & $1/1.18$ & $0.351$\\
 $30$ & $10$  & $0.621$ & $0.74$ & $2.72$ & $1/1.14$ & $0.351$\\
   \hline \hline 
\end{tabular}
  \end{center}
\end{table}

\subsection{Fitting Results for High $\dot{m}$-case}

Now, for applying to the ULXs, we examine the high $\dot{m}$-case.
Here, we apply the correction that we found in low $\dot{m}$-case
to the high $\dot{m}$-case. That is, we use $\xi^{\rm trans}=0.353$.
Note that in addition to $\kappa$ we also found necessary correction for the
inner temperature, which is $\kappa^{\rm trans}=1/1.2$, and thus the real
correction for the inner temperature is $\kappa^{\rm real}=\kappa/1.2$.
However, as we decided to treat this correction separately, the most
important basic correction for our fitting methodology is the new value
of $\xi$, which is $\xi^{\rm trans}=0.353$, while still using
$\kappa \sim 1.7$ instead of $\kappa^{\rm real}=\kappa^{\rm trans}\kappa$
for the fitting.
The detailed results of the fitting are summarized in table 4 and 5. 

\subsubsection{$m$-dependence}
Firstly, let us see the fitting results for different black hole masses.
We calculated the spectra of the slim disk for $m=10$, $30$, and $100$, 
fixed $\dot{m}$ at 100, and considered the face-on case only, $i=0$
in this subsection.
Let us focus on the inner temperature, $kT_{\rm in}$, and $p$ (figure 2). 

We see that the inner temperature, $kT_{\rm in}$, decreases as the mass 
increases. 
This is natural, since the more massive is the black hole the lower is the 
temperature, as $T_{\rm in} \propto M^{-1/4}$.
Interestingly, the value of $p$ behaves in the same way, which is not obvious. 
It is important to note that $p$ is determined by the slope in the lower 
energy (soft X-ray) region (Watarai et al. 2000), that is the middle part 
of the spectrum. The disk spectrum is shifted to the lower energy region as 
the mass increases (see Fig. 3). 
In other words, a fixed fitting energy range corresponds to a different part of
the spectrum for different black hole mass (see figure 3 for illustration). 
This is the reason why $p$ depends on the black-hole mass.

We found that the higher is the black hole mass, the smaller is the $p$-value
(see table 4).
Note that recent X-ray fitting of some ULX spectra with the extended DBB
model also showed that $p \sim 0.5$.
This is natural if we jump to conclusion that those ULXs have high black-hole
masses based on our finding above.
However, a small $p$ can also be obtained even by a stellar-mass black hole
with a high mass-accretion rate, which is discussed more in the next
subsection.
The important thing is that the dependence of $p$ on black-hole mass that we
found here is natural, since we restrict the fitting energy range. 
 
\begin{figure}
  \begin{center}
\centerline{\epsfig{file=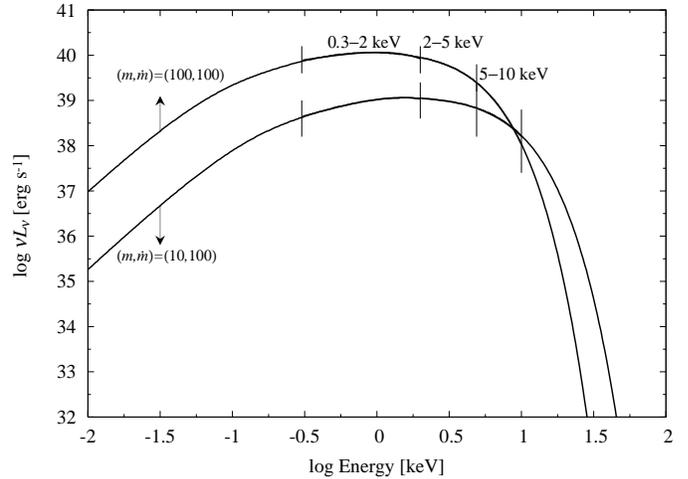,width=6.5cm,angle=270}}
  \end{center}
  \caption{Fitting range sample}\label{Figure:3}
\end{figure}
%
Now, let us see the behavior of the inner radius obtained from the fitting.
We found that the inner radius also depends on the black-hole mass;
a larger black-hole mass has a smaller $r_{\rm in}/r_{\rm g}$.
In the fitting, the innermost radius is mainly determined by the high-energy
part of the spectrum (inner region of the disk).
The dependence of the inner radius on the black-hole mass shows that the
shapes of the spectrum from the inner part of the disk change with the 
black-hole mass. In other words, the spectrum of the inner part of the 
disk of a massive black hole is not a simple scale up of that of the 
stellar-mass black hole (cf. \cite{3}).
However, the difference is small and becomes less prominent in the lower
energy part (outer part of the disk).

\subsubsection{$\dot{m}$-dependence}
Now, let us see the fitting results for various mass accretion rate,
$\dot{m}$ (table 4).
The most interesting result is the luminosity. 
In the theory of a slim disk, the luminosity at first increases as the 
mass-accretion rate increases. However, above a certain value, the radiation 
efficiency decreases as the advection effect becomes dominant, and the 
luminosity increases not at the same rate as the mass-accretion rate 
(Jaroszy\'{n}ski et al. 1980). 
In other word, the luminosity is saturated above a certain mass-accretion 
rate value.

We found that the luminosity rapidly increases as we increase $\dot{m}$ from
32 to 100.
As we further increase $\dot{m}$ up to 1000, slower increase in luminosity
is realized (figure 4).
In other words, our results are in good agreement with the theory,
and thus justifies our methodology.
Note that the fitting results of the luminosity for different black hole
masses coincide in figure 4.

As we briefly mentioned above, $p$ decreases as $\dot{m}$ increases which is
also in good agreement with theoretical prediction.
Since as $\dot{m}$ increases the heat is trapped in the flow of matter and thus
the advective, horizontal heat flux becomes important (\cite{3}).
In other words, the effective temperature gradient becomes flatter as $\dot{m}$
increases, and thus $p$ decreases.
Although higher black hole mass gives lower $p$-value, it is important
to remember that a higher black-hole mass also gives a lower $T_{\rm in}$.
Therefore, it is more likely that a lower mass black hole with higher mass
accretion rate can explain high temperature, low $p$-value ULXs.

\subsubsection{$i$-dependence}
Lastly, we would like to show the behavior of the spectra for various
inclination angles.
We mentioned earlier when we described the spectrum model of Watarai 
et al. (2005) that the spectral shape of their model strongly depends on 
the inclination angle as the mass accretion rate increases.
Let us now see the fitting results for various inclination angles. 
Here, we fixed $\dot{m}=100$, and $m=10$, $30$, and $100$ (table 5).
The inner temperature of the disk at first tends to increase as the 
inclination angle increases, but a sudden drop is realized at an inclination 
angle above $50^{\circ}$ (figure 5).
The slight increase in $kT_{\rm in}$ at low inclination angles is basically 
caused by the Doppler boosting, while a sudden drop above $50^{\circ}$ is 
caused by the self-obscuration effect.
Note that we obtained very similar trend even for different black hole masses.
In this study, we focus on the case for inclination angle below 
$50^{\circ}$ to explain ULXs with high inner temperature, as suggested 
by Watarai et al. (2005).

\subsubsection{Mass Correction Factor, $m/m_{\rm x}$}
Let us now examine the mass correction factor, which we define as the ratio
between the real black hole mass, $m$, to the estimated mass by the X-ray
spectral fitting, $m_{\rm x}$.
Here, we adopt the assumption used in the conventional disk blackbody model, 
that is $r_{\rm in}=3r_{\rm g}$.

Firstly, let us see the dependence of the correction factor on $\dot{m}$
(figure 4).
Here, we fix $i=0$ and consider three different masses:
$m=10$, $30$, and $100$.
Interestingly, we can see a similar trend in all cases;
the correction factor increases as the mass-accretion rate increases. 
Interestingly, we can see that the correction factor also seems to depend
on the
black-hole mass (see table 4). It follows a similar trend as that of 
$\dot{m}$; that is, a larger correction factor is needed for a larger
black-hole mass, although it is less prominent for the small $\dot{m}$ case.

These findings show that correction factor ($\sim 2$) is needed when we
work with the system of high $m$ and $\dot{m}$. Unfortunately, those are 
indeed two pieces of information that we intend to find. 
However, since we are interested in the high temperature ULXs,
we suggest the correction factor of 1.2 -- 1.3 and 1.3 -- 1.6 for
$m=10$ and $30$, respectively, depending on the mass-accretion rate.

Next, let us see the dependence of the correction factor on the 
inclination angle (figure 6).
We found that as long as we consider inclination angles below $50^{\circ}$, 
the correction factor does not have a strong dependence on the inclination 
angle.
Here, we show three examples for $\dot{m}=100$: $m=10$, $30$, and $m=100$. 

\begin{figure*}
  \begin{center}
\centerline{\epsfig{file=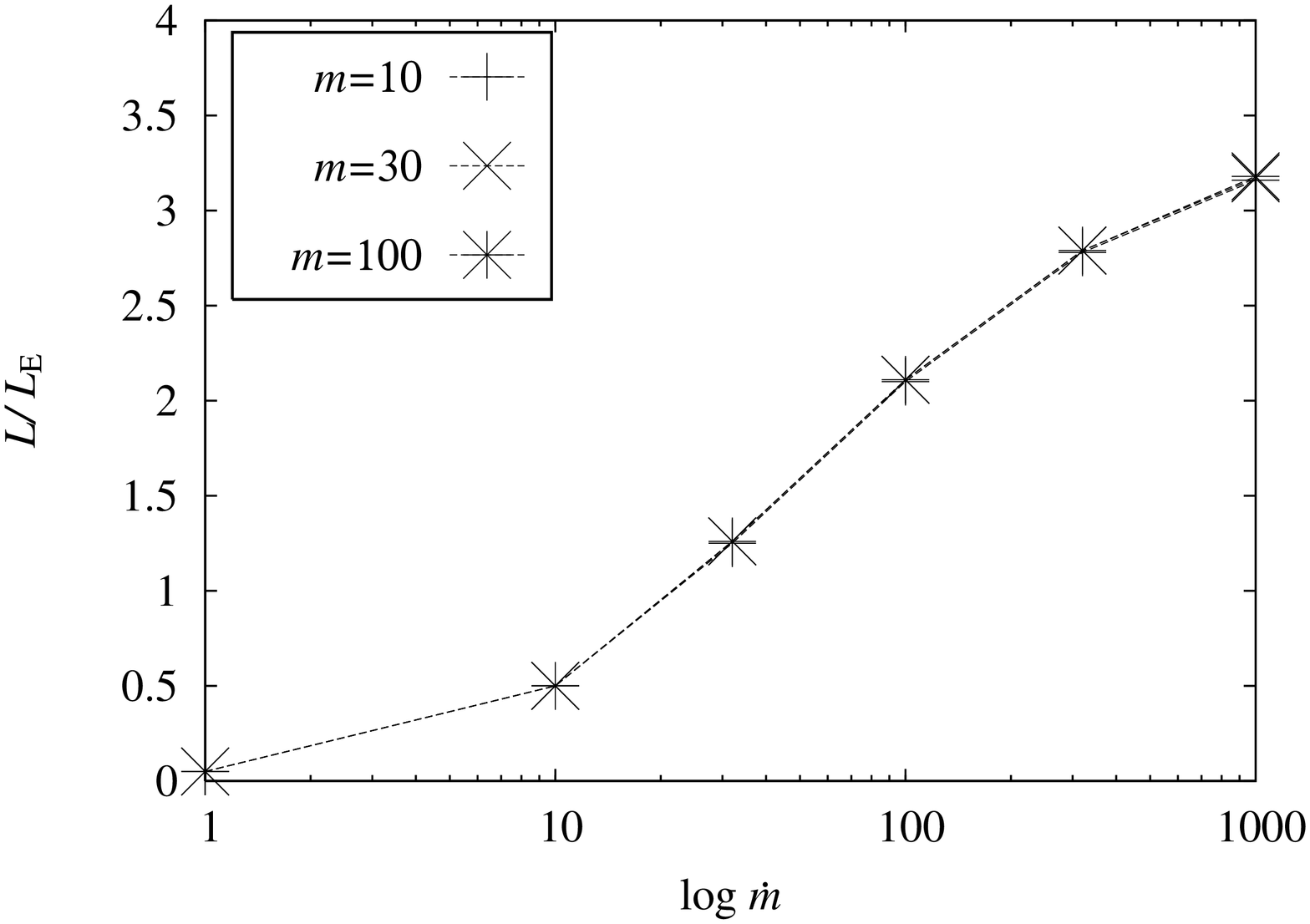,width=6.50cm,angle=270} \epsfig{file=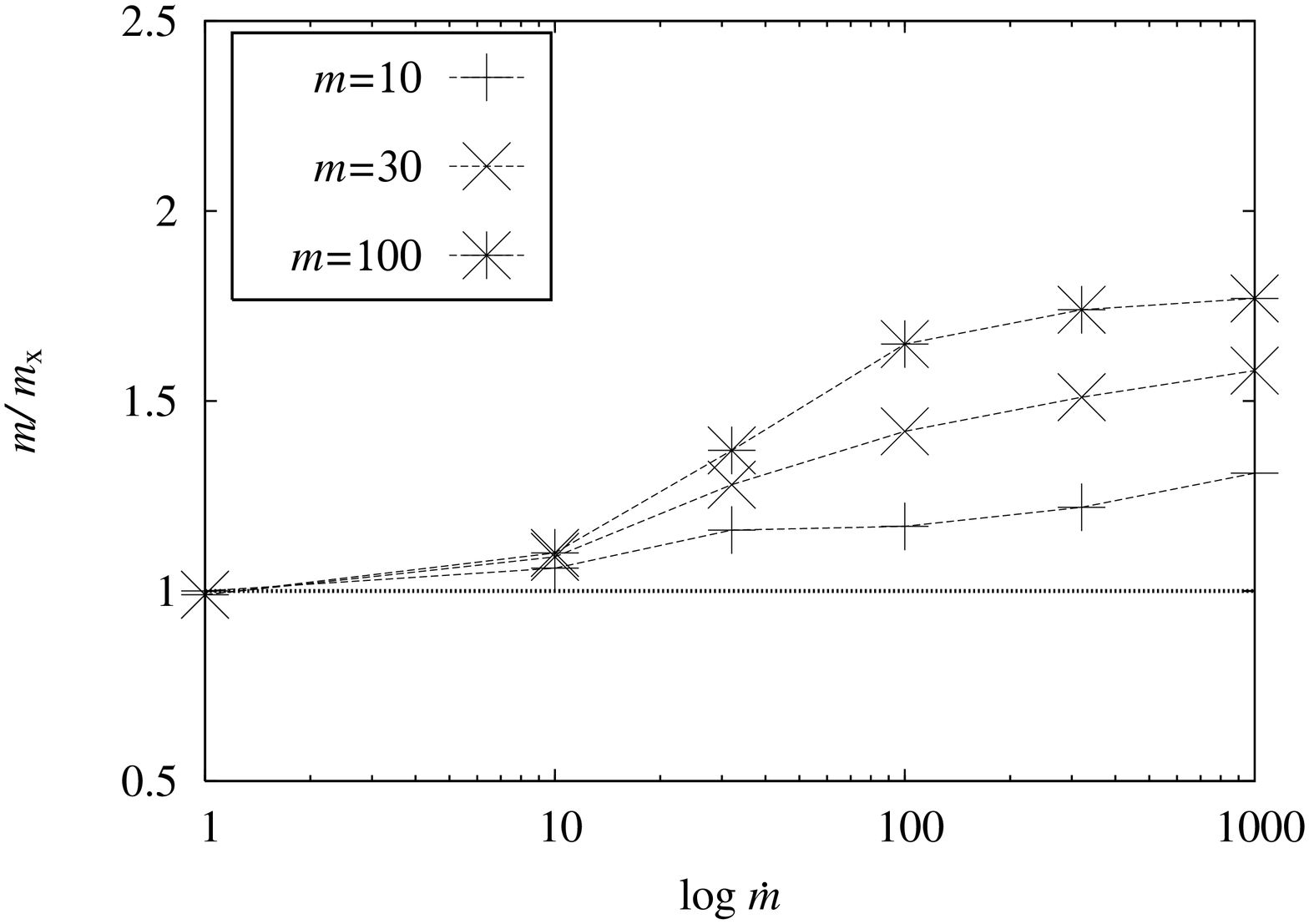,width=6.5cm,angle=270}}
  \end{center}
  \caption{Fitting results: $L/L_{\rm E}$ (left panel) with $i=0$ and $m=10,\ 30$, and the correction factor (right panel), $m/m_{\rm x}$, for various $\dot{m}$, calculated for $m=10$, $30$, and $100$ with $i=0$.}\label{Figure:4}
\end{figure*}
%

\begin{figure}
  \begin{center}
\centerline{\epsfig{file=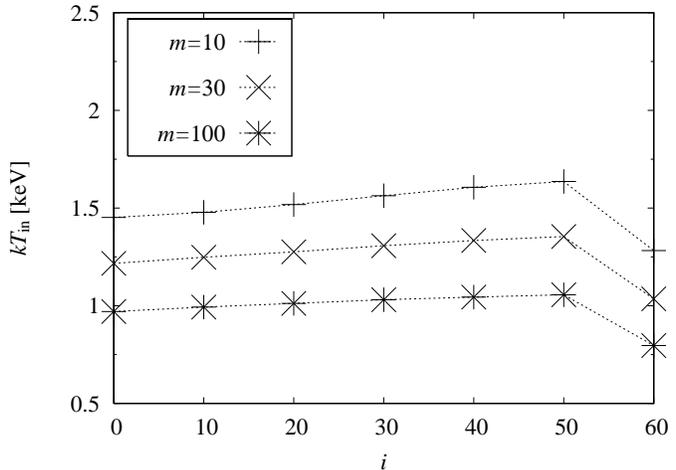,width=6.5cm,angle=270}}
  \end{center}
  \caption{Fitting results: $kT_{\rm in}$ with $m=10$, $30$, and $100$ and $\dot{m}=100$ for various $i$.}\label{Figure:4}
\end{figure}
%

\begin{figure}
  \begin{center}
\centerline{\epsfig{file=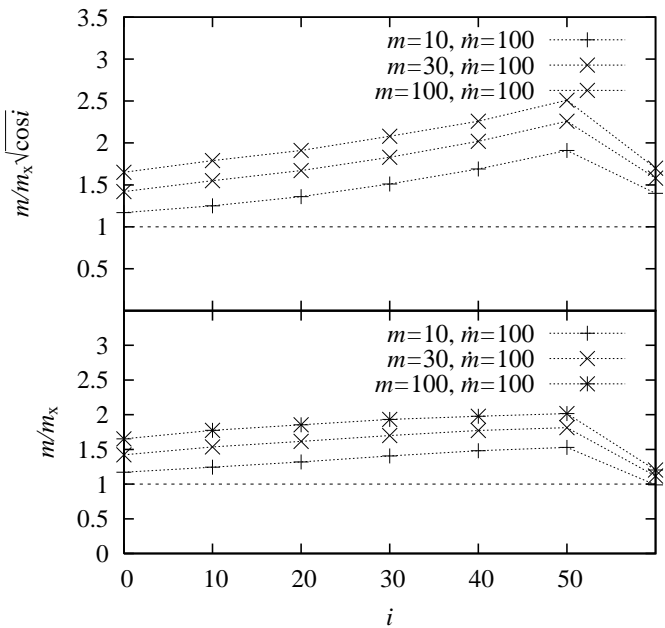,width=8.5cm,angle=270}}
  \end{center}
  \caption{Correction factor ($\equiv m/m_{\rm x}$). Top panel: for $m=10,\ 30$, and $100$ and $\dot{m}=100$ for various $i$ when $i$ is unknown. Bottom panel: same as top panel but when $i$ is known.}\label{Figure:5}
\end{figure}

\begin{table}
  \caption{Fitting results with the extended DBB model for $\alpha=0.01$ and $i=0$. $\xi^{\rm trans}=0.353$ is assumed to derive $r_{\rm in}$ and the real $kT_{\rm in}$ is the value in the table multiplied by $\kappa^{\rm trans}$. 
}\label{tab:first}
  \begin{center}
    \begin{tabular}{rrrrrrr} 
\hline \hline
 $m$ & $\dot{m}$ & $p$ & $kT_{\rm in}$ & $r_{\rm in}/r_{\rm g}$ & $m/m_{\rm x}$ & $L/L_{\rm E}$ \\
  \hline
 $10$ & $32$   & $0.605$ & $1.27$ & $2.55$ & $1.16$ & $1.25$\\
 $10$ & $100$  & $0.559$ & $1.45$ & $2.52$ & $1.17$ & $2.10$\\
 $10$ & $320$  & $0.518$ & $1.59$ & $2.41$ & $1.22$ & $2.78$\\
 $10$ & $1000$ & $0.498$ & $1.70$ & $2.25$ & $1.31$ & $3.16$\\
  \hline
 $30$ & $32$   & $0.583$ & $1.02$ & $2.30$ & $1.28$ & $1.26$\\
 $30$ & $100$  & $0.535$ & $1.22$ & $2.08$ & $1.42$ & $2.11$\\
 $30$ & $320$  & $0.500$ & $1.34$ & $1.96$ & $1.51$ & $2.79$\\
 $30$ & $1000$ & $0.484$ & $1.42$ & $1.87$ & $1.58$ & $3.17$\\
  \hline
 $100$ & $32$   & $0.563$ & $0.78$ & $2.16$ & $1.37$ & $1.26$\\
 $100$ & $100$  & $0.511$ & $0.97$ & $1.79$ & $1.65$ & $2.11$\\
 $100$ & $320$  & $0.484$ & $1.07$ & $1.70$ & $1.74$ & $2.79$\\
 $100$ & $1000$ & $0.474$ & $1.11$ & $1.67$ & $1.77$ & $3.18$\\
  \hline \hline
\end{tabular}
  \end{center}
\end{table}
%
\begin{table}
  \caption{Same as table 4 but here we fixed $\dot{m}=100$ while varying the inclination angle, $i$.
}\label{tab:first}
  \begin{center}
    \begin{tabular}{rrrrrrr} 
\hline \hline
  $m$ & $i$ & $p$ & $kT_{\rm in}$ & $r_{\rm in}/r_{\rm g}$ & $m/m_{\rm x}$ & $L/L_{\rm E}$  \\
  \hline
  $10$ & $0$  & $0.559$ & $1.45$ & $2.52$ & $1.17$ & $2.10$ \\
  $10$ & $10$ & $0.553$ & $1.48$ & $2.37$ & $1.24$ & $1.97$ \\
  $10$ & $20$ & $0.552$ & $1.52$ & $2.24$ & $1.32$ & $1.86$ \\
  $10$ & $30$ & $0.553$ & $1.56$ & $2.10$ & $1.40$ & $1.70$ \\
  $10$ & $40$ & $0.558$ & $1.60$ & $1.99$ & $1.48$ & $1.50$ \\
  $10$ & $50$ & $0.566$ & $1.64$ & $1.93$ & $1.53$ & $1.28$ \\
  $10$ & $60$ & $0.585$ & $1.28$ & $2.99$ & $0.99$ & $0.90$ \\
  \hline
  $30$ & $0$  & $0.535$ & $1.22$ & $2.08$ & $1.42$ & $2.11$ \\
  $30$ & $10$ & $0.528$ & $1.25$ & $1.92$ & $1.54$ & $1.97$ \\
  $30$ & $20$ & $0.529$ & $1.28$ & $1.83$ & $1.62$ & $1.86$ \\
  $30$ & $30$ & $0.532$ & $1.31$ & $1.74$ & $1.70$ & $1.70$ \\
  $30$ & $40$ & $0.538$ & $1.33$ & $1.67$ & $1.77$ & $1.51$ \\
  $30$ & $50$ & $0.546$ & $1.35$ & $1.63$ & $1.81$ & $1.28$ \\
  $30$ & $60$ & $0.562$ & $1.04$ & $2.65$ & $1.12$ & $0.90$ \\
  \hline
  $100$ & $0$  & $0.511$ & $0.97$ & $1.79$ & $1.65$ & $2.11$ \\
  $100$ & $10$ & $0.506$ & $0.99$ & $1.66$ & $1.78$ & $1.98$ \\
  $100$ & $20$ & $0.508$ & $1.01$ & $1.59$ & $1.85$ & $1.86$ \\
  $100$ & $30$ & $0.512$ & $1.03$ & $1.53$ & $1.93$ & $1.70$ \\
  $100$ & $40$ & $0.520$ & $1.04$ & $1.49$ & $1.98$ & $1.51$ \\
  $100$ & $50$ & $0.528$ & $1.06$ & $1.47$ & $2.02$ & $1.28$ \\
  $100$ & $60$ & $0.541$ & $0.80$ & $2.46$ & $1.20$ & $0.90$ \\
   \hline \hline 
\end{tabular}
  \end{center}
\end{table}

%

To summarize, for high $\dot{m}$-case, we need a mass correction factor
defined as $m/m_{\rm x}$.

Before leaving this section, it is important to note that in calculating the
disk structure, and thus the spectrum, we set the inner calculation boundary
at $2r_{\rm g}$.
On the other hand, to derived the black hole mass we adopt
$r_{\rm in}=3r_{\rm g}$.
Since the correction factor that we found here varies with $m$ and also 
$\dot{m}$, we conclude that the inner radius obtained from the fitting does 
not represent the inner edge of the disk as was stressed in Watarai and 
Mineshige (2003a).
They suggest that the inner radius from the fitting represents the radiation
edge outside which substantial emission is produced.
In fact, for super-critical accretion disks, the emission produced inside
$3r_{\rm g}$ is really significant (Watarai et al. 2000).  

\subsection{X-ray HR Diagram}

Finally, we plot our results in the so-called X-ray HR diagram, which shows 
the relation of X-ray luminosity, $L_{\rm x}$, with the inner temperature
(in keV), $kT_{\rm in}$, obtained from the fitting (figure 7).
Here, we now include the correction for $T_{\rm in}$ in the X-ray HR diagram.
That is $T_{\rm in}=\kappa T_{\rm eff}/1.2$

We plot for $m=10$ (in black \lq $+$\rq ), $30$ (in black \lq $\times$\rq ),
and $100$ (in black \lq $\ast$\rq ) and $\dot{m}=1,\ 10,\ 32$, $100$, $320$,
and $1000$, while $i=0$.
This figure is similar to that of Watarai et al. (2005). However, in their
paper, they only calculated for $m=10$, and then interpolated the results for 
higher black-hole masses.
We also have included the required corrections discussed in subsection
3.1.

For low $\dot{m}$, our results follow $L \propto T_{\rm in}^{4}$, as expected
from the sub-critical accretion (standard accretion disk) theory.
For high $\dot{m}$ our results follow $L \propto T_{\rm in}^{2}$.
We also plot some ULXs and well-known BHCs by using the references in 
Vierdayanti et al. (2006).
For BHC whose inclination angle has been known, we included it in
calculating the luminosity, while for those whose inclination angle is unknown,
we assumed a face-on case.
Some ULXs can very likely be explained as stellar-mass black holes
with a super-critical mass accretion rate.
Except for the source IC 342 S1, Vierdayanti et al. (2006) fitted the other
ULX data with the extended DBB model.
The data of IC 342 S1 are taken from Mizuno et al. (2001), in which they fitted
the ASCA data with the multicolor disk blackbody (DBB) model.

The red dotted lines shows the masses and $\eta$ ($\equiv L/L_{\rm E}$)
which were calculated from the standard disk relations [see \cite{25} equation
(9) and (11)]. 
We also plot the fitting results from Shakura and Sunyaev analytical solutions
(see section 3.1) in red \lq $\otimes$\rq mark.
We can thus safely conclude that the black holes in ULXs, which Vierdayanti
et al. (2006) analyzed, are stellar-mass black holes with mass $< 100M_{\odot}$.

\begin{figure*}
  \begin{center}
\centerline{\epsfig{file=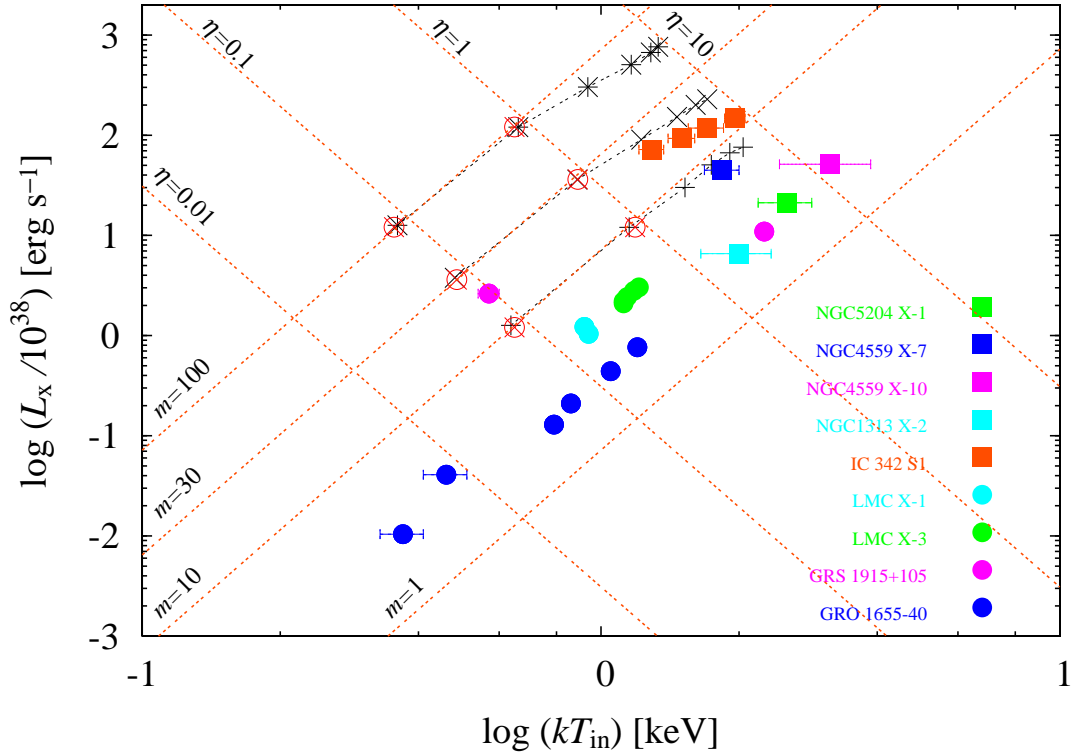,width=10.50cm,angle=270}}
  \end{center}
  \caption{Luminosity vs. temperature diagram for various $m$ ($m=10$
  in black \lq $+$\rq, $m=30$ in black \lq $\times$\rq, and $m=100$ in
  black \lq $\ast$\rq ) and $\dot{m}=1,\ 10,\ 32$, $100$, $320$, and
  $1000$ (increasing with luminosity and $kT_{\rm in}$), while
  $i=0$. Several ULXs (NGC 5204 X-1, NGC 4559 X-7, NGC 4559 X-10, NGC
  1313 X-1, and IC 342 Source 1) and some well-known black hole candidates (LMC X-1, LMC X-3, GRO 1655-40, and GRS 1915+105) are included in the diagram (reference:Vierdayanti et al. 2006, Watarai et al. 2005). Red \lq $\otimes$\rq symbols are calculated from analytical solution of standard disk and red lines are calculated from standard disk relations (see text, section 3.3).}\label{Figure:4}
\end{figure*}
%
%

\section{Discussion}

In this section we would like to discuss some important issues that
are related to our present study.
The first and the most important issue is the dependence of the fitting
results on the fitting energy range.
The next two issues are the spectral hardening factor, which remains an open
question for the case of super-critical accretion, and
outflow of material from the disk.
Another interesting issue is the formation scenario of ULX in terms of binary
star evolution.
Supposing some ULXs are highly accreting stellar-mass black holes, it is very
natural to think that those ULXs are binary systems and the disk material is
supplied by their companions.
In the last part of this section, we comment on other plausible black hole mass
estimation methods for ULXs.

\subsection{Dependence on the Fitting Energy Range}
It is very important to note that the fitting results sensitively depend
on the fitting energy range.
In our present study, we fixed the fitting energy range to be 0.3 -- 10 keV
which is normally used in the study of BHBs as well as ULXs.
Therefore, it is interesting to know how the fitting results will change
if we choose a different fitting energy range.
It will be necessary at some later time when the observations could
provide us with data of a wider energy range.
We extend our results to the higher energy part of the spectrum, so that we
can also predict some features that had not been
made available much by observations.
On the other hand, we do not consider the lower energy part, lower than
0.3 keV, since there remains some difficulties in the analysis of this energy
part due to the effect of the interstellar medium.

We, therefore, performed the same procedure as what we explain in section 3
for a larger fitting energy range, 0.3 -- 30 keV.
We summarize the fitting results in table 6.
We obtained higher $T_{\rm in}$ and lower $r_{\rm in}$; as a consequence
we have larger mass correction factor, ranging from 1.4 -- 1.9 ($m=10$),
1.4 -- 2.0 ($m=30$), and 1.4 -- 2.1 ($m=100$).
To conclude our discussion in this subsection, we need a mass correction,
$m/m_{\rm x} \sim 2$ at most, for fitting with a wider energy range, which can
be use as a reference when the observational data are available.

\subsection{Spectral Hardening}
The spectral hardening factor plays an important role in estimations of
the black-hole mass, particularly when it is based on the X-ray spectral
fitting.
As mentioned in section 3, the spectral hardening factor is usually set to
be 1.7 for the standard accretion disk.
This value was obtained from a numerical calculation of the vertical structure 
and the radiative transfer of an accretion disk around a Schwarzschild black
hole by Shimura and Takahara (1995), in which a geometrically thin disk
assumption is used. Their results imply that the X-ray spectrum of a
standard disk is affected by Comptonization, and the local flux can be
approximated by a diluted blackbody flux (see subsection 2.2), with
$\kappa \sim 1.7$ (see also Czerny \& Elvis 1987; \cite{58}). 

We are aware that the bound-free opacity may still play an important
role, and that it is important to address this issue properly.
In our present study, however, we did not include the effect of bound-free
opacity, and we simply mimicked the deviation from the blackbody spectra by
a diluted blackbody approximation without considering the details of the
physical mechanism that may produce it.
In fact, we also think that it is even possible that the spectrum at
high-energy part may experience softening as well (\cite{7171};
Suleimanov \& Poutanen 2006).
In order to address these issues properly, we need to work with
radiative transfer, which is beyond the scope of our present study.
Regarding the effect of the bound-free opacity on the spectral shape,
it will be necessary for the readers to consult other papers
(e.g., Suleimanov \& Poutanen 2006).
Meanwhile, we continue our discussion in the context of the diluted blackbody
approximation in which spectral hardening factor plays a key role.

Merloni et al. (2000) analyzed the spectral hardening factor used to correct
the fitting results of the DBB model. They use a self-consistent model for
the radiative transfer and the vertical temperature structure in a standard
disk to simulate the observed disk spectra.
In addition, they also take into account the gravitational redshift
and transverse Doppler effects. 
They fitted this model to the DBB model, and found that the spectral hardening
factor is not constant; the spectral hardening factor becomes higher when the
accretion rate and/or coronal activity is high. The range 
of the varying hardening factor is within $1.7<\kappa<3$.
Davis et al. (2006) also show significant spectral hardening when a disk 
becomes effectively optically thin.

In terms of super-critical accretion model, Kawaguchi (2003) showed that
spectral hardening by electron scattering is quite $\alpha$ sensitive, since
$\tau_{\rm es}/\tau_{\rm abs} \propto \alpha^{1/8}$. Therefore, an
increase of $\alpha$ results in an increase of the color temperature, and thus
we get more spectral hardening.
Note that all the spectral calculations mentioned above assume hydrostatic
disk layers.

In the present study, we set the spectral hardening factor to be,
$\kappa=1.7$, the widely used value for the case of sub-critical (standard)
accretion due to the uncertainty of the $\kappa$ in the case of super-critical
accretion.
The choice of other value of $\kappa$ would certainly affect the spectrum
shape, for instance, choosing $\kappa>1.7$ will cause the spectrum to be
harder in the high energy band and the slope in the lower energy band would
also be changed. $\kappa\sim 3$ was suggested for high luminosity disks
(Watarai \& Mineshige 2003b).
Since the derived black-hole mass from the fitting depends on
$\kappa$ as $M \propto \kappa^{2}$, a mass
underestimation/overestimation is possible due to the choice of $\kappa$.

To summarize, there remain uncertainties in the spectral hardening factor
of super-critical accretion flow.
This will, in turn, introduce another source of uncertainty in our present
mass estimation.
A more complete description of spectral properties of super-critical
accretion is awaited.

\subsection{Outflow of Material}
Through a 2-dimensional radiation-hydrodynamic simulation Ohsuga et al.
(2005) found high-speed outflow (with 10--20 \% of speed of light)
from the center of super-critical flow with a rate of $\sim L_{\rm E}/c^2$.
This nearly relativistic motion will affect the photon spectra, whose effect
depends on the inclination angle of the system (Heinzeller et al. 2006).

In fact, Heinzeller et al. (2006) found that
the average photon energy due to a shift of the
frequency of the escaping photons by the relativistic Doppler effect
in a face-on case (small inclination angle) is of a factor of 1.18
higher than that of the edge-on case (large inclination angle).
They also reported that the main features of the thermal component of
the spectrum of super-critical accretion flow are consistent with
the slim disk, while there remain unresolved features in the high-energy
part.
This kind of simulation is very important to address the effect
of the outflow in the observed spectrum. However, an adequate calculation
resolution has not been available at the moment.

In the meantime, as a qualitative study, we compare two different cases.
The first case is a disk with advection without outflow (our case),
and the second case is the disk with outflow without advection, which was
first considered by Shakura and Sunyaev (1973), as also discussed in Poutanen
et al. (2007).
In addition to the fact that both cases result in similar temperature
profiles, $T_{\rm eff} \propto r^{-1/2}$ (flatter temperature distribution
than that of the standard disk), we found another interesting fact.

Both cases have a characteristic radius.
In the advection case, it is called the trapping radius, $r_{\rm tr}$,
inside of which the diffusion timescale is longer than the accretion timescale
(and thus the photon trapping becomes important).
In the outflow case, it is called the spherization radius, $r_{\rm sp}$, at
which the disk luminosity is close to the Eddington luminosity (\cite{7474}).
Interestingly, those two radii are comparable or, to be more precise, the
spherization radius is twice as small as the trapping radius:
$r_{\rm tr} \sim \dot{m} r_{\rm g}/2$ while
$r_{\rm sp} \sim \dot{m} r_{\rm g}/4$, if we set $H/r \sim 1$ for the
advection case.
This means that the advection effect becomes effective even before the outflow
needs to set in to remove any excess of the gravitational energy caused by
high mass-accretion rate inflow of the disk materials.

Although it is necessary to address this issue more properly,
especially when both advection and outflow occur simultaneously
(see also \cite{34}), we did not include the effect of the outflow in our
present study, while assuming that the advection works effectively.

\subsection{On Black-Hole Binary Evolution}

The work in the disk model is indeed synchronized by the work in the field of
black-hole evolution in binary systems.
We point out some studies on black-hole binary evolution in this section.

King et al. (2001), for example, proposed a mild X-ray beaming model for ULXs.
Their model suggest that ULXs may represent the phase of thermal-timescale
mass transfer that occurs in high-mass X-ray binaries.
This phase is short-lived, but is extremely common in the evolution of X-ray
binaries.
The short lifetimes of high-mass X-ray binaries could explain the association
of ULXs with the star-formation region.
Moreover, in the context of a binary system, their model does not
require IMBH to explain the nature of ULXs.

Rappaport et al. (2005) suggest that the luminosity of ULXs can be explained
by a stellar-mass black hole in a binary with a moderate mass companion.
They follow the evolution of the binary before the black hole is formed until
it emits the X-ray by the accretion process onto the black hole.
They found that luminosities of the order of ULX luminosities can be produced
if the Eddington limit is allowed to be exceeded by about a factor of ten.
Interestingly, this clearly supports super-critical accretion flow onto
a black hole as a possible mechanism for ULXs.

\subsection{Other Black Hole Mass Estimation Methods}
Despite being very useful, some uncertainties remain concerning mass
estimation through the X-ray spectral fitting as a result of its attachment
on some assumptions.
Therefore, mass estimation from other independent methods are eagerly awaited
to settle the dispute over the black-hole mass of the ULXs.
It also remains possible that ULXs cannot be explained by a single
theoretical model.

Despite being widely used when the dynamical mass estimation method cannot
be applied, the X-ray spectral fitting is not the only method
to estimate the black-hole mass. Indeed, other independent mass estimation is
demanded to settle the dispute over the black hole mass of the ULXs.
One of the other independent mass estimations is by double peak
quasi-periodic oscillation (QPOs) in a 3:2 ratio to be discovered in the time
variability of ULXs (\cite{5}).
These QPO frequencies scale inversely with the mass of the source,
which was first shown for microquasars by McClintock and Remillard (2003),
and more recently for low-mass Seyfert galaxies by Lachowicz et al. (2006).
For ULXs, in general, it has not been detected yet.
Recently, Strohmayer et al. (2007) reported possible double peak QPOs in
NGC 5408 X-1 which happened to have 4:3 ratio. Unfortunately, the
scaling law of the upper frequency QPO and the mass is not obvious for
QPOs other than that with a 3:2 ratio.
 
Other scaling laws are found in X-ray binaries and active galactic nuclei
(AGNs).
It was found that the ratio of the radio to X-ray emission depends strongly
on the black hole mass (Falcke et al. 2004).
If ULXs are in the same family with our Galactic X-ray binaries in the very
high state, it is expected that they could also be radio transients.
In this regard, K\"{o}rding et al. (2004) conducted a radio-monitoring
study of ULXs in nearby galaxies with the VLA.
K\"{o}rding et al. (2005) found that generally ULXs do not emit steady-state
radio emission above radio powers of $1.5 \times 10^{17}$ W/Hz, and thus
radio loud ULXs is still awaited to be found.

Pooley and Rappaport (2005) suggest that X-ray and optical eclipses can
be used to distinguish the two possible black hole masses.
ULXs with stellar-mass black holes should exhibit at least twice as many
eclipses as would IMBH systems.
They proposed X-ray monitoring campaigns for a few days to detect eclipses.
In the optical range, monitoring $\sim 10$ ULXs twice a night each for a
couple of months by using ground-based telescope should also work.

Meanwhile, we can merely rely on X-ray spectral fitting which was
found to be quite a reliable method in the case of Galactic black-hole
binaries.

\begin{table}
  \caption{Same as table 4 but the energy range for fitting is 0.3 -- 30 keV. 
}\label{tab:first}
  \begin{center}
    \begin{tabular}{rrrrrrr}
\hline \hline
 $m$ & $\dot{m}$ & $p$ & $kT_{\rm in}$ & $r_{\rm in}/r_{\rm g}$ & $m/m_{\rm x}$ & $L/L_{\rm E}$ \\
  \hline
 $10$ & $32$   & $0.58$ & $1.39$ & $2.12$ & $1.39$ & $1.25$\\ 
 $10$ & $100$  & $0.53$ & $1.79$ & $1.66$ & $1.78$ & $2.10$ \\
 $10$ & $320$  & $0.50$ & $1.96$ & $1.59$ & $1.86$ & $2.78$ \\
 $10$ & $1000$ & $0.48$ & $2.04$ & $1.57$ & $1.88$ & $3.16$ \\
  \hline
 $30$ & $32$   & $0.57$ & $1.07$ & $2.08$ & $1.42$ & $1.26$\\ 
 $30$ & $100$  & $0.51$ & $1.41$ & $1.56$ & $1.90$ & $2.11$ \\
 $30$ & $320$  & $0.48$ & $1.54$ & $1.49$ & $1.99$ & $2.79$ \\
 $30$ & $1000$ & $0.47$ & $1.60$ & $1.48$ & $2.00$ & $3.17$ \\
  \hline
 $100$ & $32$   & $0.55$ & $0.80$ & $2.06$ & $1.40$ & $1.26$\\ 
 $100$ & $100$  & $0.49$ & $1.07$ & $1.48$ & $2.00$ & $2.11$\\
 $100$ & $320$  & $0.47$ & $1.17$ & $1.41$ & $2.09$ & $2.79$ \\
 $100$ & $1000$ & $0.46$ & $1.21$ & $1.41$ & $2.09$ & $3.18$ \\
   \hline \hline 
\end{tabular}
  \end{center}
\end{table}

\section{Conclusion}

We have analyzed the assumption $r_{\rm in}=3r_{\rm g}$ for black hole mass
estimation for the case of super-critical accretion disk.

We fitted the slim disk spectra for various parameter combinations
($m$, $\dot{m}$, and $i$), with the extended disk blackbody model.
We fixed $\alpha=0.01$ in the disk structure calculation, and chose
$\kappa=1.7$ for the spectrum calculation.
To estimate the black hole mass, we adopt $\xi^{\rm trans}=0.353$.

As expected from a theoretical point of view, we obtained $p$-values that
deviate from 0.75, the value expected for the standard disk model.
Moreover, we found that by adopting, the assumption $r_{\rm in}=3r_{\rm g}$
for a slim disk, the estimated black hole mass should be corrected.
The correction factors are 1.2 -- 1.3 and 1.3 -- 1.6 for $m=10$ and $m=30$,
respectively.
Since the corrections are small, we can safely conclude that the black
holes in ULXs which Vierdayanti et al. (2006) analyzed are stellar-mass
black holes with mass $< 100M_{\odot}$.

\bigskip
\bigskip
We gratefully thank the anonymous referees for their useful comments and
suggestions concerning the first manuscript of this paper.
This work was supported in part by the Grants-in-Aid of the Ministry
of Education, Science, Culture, and Sport  (14079205, 16340057 S.M.;
16004706 K.W.), by the Grant-in-Aid for the 21st Century COE 
``Center for Diversity and Universality in Physics'' from the Ministry
of Education, Culture, Sports, Science and Technology (MEXT) of Japan.
One of the authors, K.V., gratefully thank the MEXT scholarship.

\end{document}